\documentclass[12pt]{article}

\newcommand{\myskip}{\vspace{\baselineskip}}
\newcommand{\mysection}[1]{\par\myskip\noindent\textbf{#1}\myskip\par}

\begin{document}
\begin{center}
{\large\textbf{Magnetoexcitons in quantum-ring structures: a novel
magnetic interference effect}}

\myskip S.~E.~Ulloa$^a$, A.~O.~Govorov$^{a,b}$,
A.~V.~Kalameitsev$^b$, R.~Warburton$^c$, and K.~Karrai$^d$\\~~ \\
$^a$ Department of Physics and Astronomy, and CMSS Program, Ohio
University, \\ Athens, OH 45701, USA; ulloa@helios.phy.ohiou.edu
\\ $^b$Institute of Semiconductor Physics, RAS,
Siberian Branch, 630090 Novosibirsk, Russia
\\
$^c$ Department of Physics, Heriot-Watt University, Edinburgh, UK
\\ $^{d}$  Sektion Physik der Ludwig-Maximilians-Universit\"at and
Center for Nano-Science, Geschwister-Scholl-Platz 1, 80539
M\"{u}nchen, Germany

\myskip \mysection{Abstract}

\myskip
\parbox{4.5in}{A novel magnetic interference effect is proposed for
a neutral, but polarizable exciton in a quantum ring with a finite
width. The magnetic interference effect originates from the
nonzero dipole moment in the exciton. The ground state of exciton
acquires a nonzero angular momentum with increasing normal
magnetic field. This leads to the suppression of the
photoluminescence in defined windows of the magnetic field.}
\end{center}
\vskip 0.4cm

 When a quantum particle moves along a closed
trajectory in external electric and magnetic fields, the
Aharonov-Bohm and Aharonov-Casher effects can occur, caused by
quantum interference between paths with different phases. As is
well known, the Aharonov-Bohm (AB) effect is related to a charged
particle trajectory enclosing a magnetic flux. Here we present a
novel magnetic interference effect for a {\em neutral}, but
polarizable quasi-particle. In particular, we show here
theoretically that the wave function of a neutral polarizable
exciton acquires a nonzero phase when it moves in a quantum ring
pierced by a magnetic flux. The neutral exciton wave function
becomes sensitive to the magnetic flux because of a net radial
electric-dipole moment induced by asymmetries in the nanostructure
potential. The transition to a finite phase and corresponding
orbital momentum in the polarized exciton state strongly changes
the photoluminescence (PL) spectrum of the system due to the
optical selection rules for interband transitions.

We demonstrate this novel magneto-interference effect using a
model of InAs self-organized quantum rings (QR's) \cite{rings}.
The calculated single-particle wave functions of electrons and
holes are peaked at different radii due to the potential
asymmetries and the difference in effective mass of the particles;
meanwhile, their mutual interaction correlates their motion around
the ring. The dipole polarization of the exciton can be strongly
enhanced due to a point/impurity charge in the ring center, or by
a voltage applied to a metal nano-gate in the middle of a ring. In
that case, the potential minimum for the electron is shifted from
that for the hole, and the ground state of the exciton acquires a
finite polarization.  To qualitatively demonstrate the effect, we
use a model of two nested one-dimensional (1D) rings with
different radii, one for the electron and one for the hole (insert
of Fig.~1). For rings with relatively large radii, the motion of
particles is strongly correlated, whereas, in a small system, the
state is nearly single-particle--like, as the relative weight of
the Coulomb interaction decreases with respect to the confinement
energies.

Since the vertical size of a QR is typically much smaller than the
lateral one, we will discuss only the in-plane motion. The
electron and hole in-plane potentials are approximated by
$U_{e(h)}(\rho)=m_{e(h)}\Omega^2_{e(h)}(\rho-R_{e(h)})^2/2$, where
$\rho$ is the in-plane distance to the ring center, and
$m_{e(h)}$, $\Omega_{e(h)}$, and $R_{e(h)}$ are  effective masses,
characteristic frequencies, and ring radius, respectively; the
indices $e$ and $h$ indicate the electron and hole quantities. In
the vertical, $z$-direction, the motion is strongly quantized.

In a magnetic field, the Hamiltonian of an exciton confined in a
quantum ring reads $\hat{H}=\hat{T}_e+\hat{T}_h+U_{e}+U_{h}+
U_{C}(|{\bf r}_e-{\bf r}_h|)$, where ${\bf r}_{e(h)}$ are the
in-plane coordinates, $\hat{T}_{e(h)}$ are the kinetic energies in
the presence of a normal magnetic field, and $U_{C}$ is the
Coulomb potential. Now we assume that the quantization in the
radial direction is stronger than that in the azimuthal direction.
It allows us to separate variables in the wave function,
$\Psi({\bf r}_e,{\bf
r}_h)=f_e(\rho_e)f_h(\rho_h)\psi(\phi_e,\phi_h)$. Here ${\bf
r}=(\rho,\phi)$. The radial wave functions $f_{e(h)}$ are strongly
localized near the radii, $R_{e(h)}$. The Hamiltonian describing
the wave function $\psi(\phi_e,\phi_h)$  is (up to a
$B$-independent constant term),

\begin{equation}
\hat{h}=-\frac{\hbar^2}{2m_eR_e^2}\frac{\partial^2}{\partial\phi_e^2}-
\frac{i\hbar\omega_e}{2}\frac{\partial}{\partial\phi_e}
-\frac{\hbar^2}{2m_hR_h^2}\frac{\partial^2}{\partial\phi_h^2}+
\frac{i\hbar\omega_h}{2}\frac{\partial}{\partial\phi_h}+
\frac{m_e\omega_e^2R_e^2+m_h\omega_h^2R_h^2}{8}+
u_{C}(|\phi_e-\phi_h|), \label{h}
\end{equation}
where $\omega_{e(h)}=|e|B/[m_{e(h)}c]$ are the cyclotron
frequencies of the particles, $B$ is the normal magnetic field,
and $u_c$ is the Coulomb potential averaged over the coordinate
$\rho$ involving the radial wave functions. By introducing new
variables, we can rewrite Eq.\ \ref{h} as
$\hat{h}=\hat{h}_0(\phi_0)+\hat{h}_1(\Delta\phi)$, where
$\Delta\phi=\phi_e-\phi_h$, $\phi_0=(a\phi_e+b\phi_h)/(a+b)$, and
$a=m_eR_e^2$ and $b=m_hR_h^2$. Then, the eigenfunctions and
eigenvalues can be found in the form of
$\psi(\phi_e,\phi_h)=\psi_0(\phi_0)\psi_1(\Delta\phi)$ and
$E=E_0+E_1$, respectively.  Here, the $\hat{h}_0$ operator is
given by
 $%
\hat{h}_0(\phi_0)= \varepsilon_0 \left[
-i\frac{\partial}{\partial\phi_0} +\frac{\Phi_{\Delta R }}{\Phi_0}
 \right]^2 
$, 
where $M=(m_eR_e^2+m_hR_e^2)/R_0^2$, $R_0=(R_e+R_h)/2$,
$\varepsilon_0=\hbar^2/(2R_0^2M)$, $\Phi_0=hc/e$, and
$\Phi_{\Delta R}=\pi(R_e^2-R_h^2)B=2\pi\Delta R R_0B$ is the
magnetic flux penetrating the area between the electron and hole
trajectories (insert of Fig.~1); and $\Delta R=R_e-R_h$. The
eigenvalues of $\hat{h}_0$ are
$E_0(l)=\varepsilon_0[l+\Phi_{\Delta R }/\Phi_0]^2$, where $l$ is
an integer which represents the total angular momentum of the
exciton.

The relative motion in the exciton is described by the operator
$\hat{h}_1(\Delta\phi)$, which involves the Coulomb potential. The
limit of strong Coulomb interaction implies the condition $R_0\gg
a_0^*$, where $a_0^*$ is the exciton Bohr radius. In this limit,
the wave function $\psi_1(\Delta\phi)$ is strongly localized near
the point $\Delta\phi=0$ and the ground-state energy for $l=0$ can
be written as $E_1(n=0)=E_{b}-2V\cos[2\pi\Phi_{eff}/\Phi_0]$,
where $V$ is the amplitude of tunneling from $\Delta\phi=0$ to
$\Delta\phi=2\pi$, $\Phi_{eff}\simeq\pi R_0^2B$, and $E_b$ is the
energy of a state localized near the angle $\Delta\phi=0$; $n$ is
the index of a quantum state \cite{theory}. In the
strong-interaction limit, the tunneling amplitude $V$ becomes
exponentially small, and the magnetic field dispersion of the
exciton energy comes mostly from the motion of a dipole. The
lowest energy  branches are $E(l,n=0)=E_1(0)+E_0(l)\simeq
E_b+E_0(l)$.

In the opposite case of $R_0\ll a_0^*$, we can neglect the Coulomb
interaction and solve Eq.~\ref{h} using the original variables.
The energy spectrum reads

\begin{equation}
 E(l_e,l_h)=\frac{\hbar^2}{2m_eR_e^2} \left[l_e
+\frac{\Phi_e}{\Phi_0} \right]^2+ \frac{\hbar^2}{2m_hR_h^2} \left[
l_h-\frac{\Phi_h}{\Phi_0} \right]^2, \label{eh}
\end{equation}
where $\Phi_{e(h)}=\pi R_{e(h)}^2B$, and $l_{e(h)}$ are electron
(hole) angular momenta.

In Figs.~1 and 2 we show the energy spectra and the PL intensity
of an exciton in a QR (notice that $E_{exc} = E_{gap} + E$, where
$E_{gap}$ is the semiconductor gap energy). For the
strong-Coulomb-interaction limit (Fig.~1), we show the lowest
states with $n=0$ and $l=0,\pm1,\pm2,...$, where $l$ is the total
angular momentum of the exciton. We see that the ground-state
momentum changes with increasing magnetic field from $l=0$ to
$l=-1,-2,...$, according to the equation for $E_0(l)$. This arises
as the electron and hole acquire different magnetic phases when
they move along different closed trajectories (insert of Fig.~1).
In the case of weak Coulomb interaction, the character of
ground-state transitions  in a magnetic field is a bit more
complicated. The ground state $(l_e,l_h)=(0,0)$ changes to the
states $(-1,0)$, $(-1,+1)$, $(-2,+1)$, etc., as the field
increases. The total momentum of the ground state, $l=l_e+l_h$,
changes correspondingly.

According to the selection rules for optical transitions between
the conduction and valence bands, only the zero-momentum excitons
can emit a photon. At low temperatures, a photo-generated exciton
relaxes within a short time to its ground state. Thus, in many
cases the PL spectrum demonstrates mostly the line related to the
ground state of an exciton. With increasing magnetic field, the
exciton in its ground state acquires nonzero momentum and can not
longer radiate. The darkness of the exciton in the ground state is
seen as a suppression of the calculated PL in the magnetic-field
intervals with the ground-state total momentum $l=l_e+l_h\neq0$.
The PL intensity shown in Fig.\ 2 was calculated from the relation
$I_{PL}(B)\propto P(l=0,T)$, where $P(l=0,T)$ is the probability
to find an exciton in the states with $l=0$ at finite temperature
$T$ \cite{A}. For the spacing $\Delta R$ we have chosen $\sim
20-30$ \AA.  As an example, an impurity charge $+|e|$ in the
middle of a ring having $R_0=85$ \AA, $\hbar\Omega_e=35 meV$, and
$\hbar\Omega_h=25 meV$, induces a shift $\Delta R\sim 20$\AA. A
similar $\Delta R$ is obtained for the ring parameters of the
system studied in Ref.\ \cite{rings}.


A similar AB effect for a neutral exciton can occur in
(spatially-indirect) type-II quantum dot embedded in a 2D quantum
well \cite{Kalameitsev}. In such a system, the electron can move
in a quantum-ring potential due to the joint action of the Coulomb
force and the quantum-dot potential. This type-II geometry would
correspond to that of GaSb/GaAs quantum dots, for example
\cite{Hatami}.  Similarly, confinement asymmetries in a ring
arising from mass differences or effective potential profiles
could also give rise to a finite polarization of the exciton
ground state.  The details of the system would determine the
strength of the magnetic field sensitivity such as shown in the
figures.

It is important to emphasize that the predicted effect depends on
the magnetic flux $\Phi_{\Delta R}$ through the area {\em between}
the electron and hole trajectories (insert of Fig.\ 1) and does
{\em not} include an exponentially-small factor due to
electron-to-hole tunneling along the ring. This is in contrast to
the AB effect for excitons in a 1D ring described recently in the
literature \cite{theory}.  This difference would make the
experimental detection of the effect discussed here much more
likely. As single-dot spectroscopy \cite{ringsPL} permits one to
observe the PL energies with very high accuracy, it would be a
suitable method to study the predicted magnetic-field interference
effects.

{\bf Acknowledgements}.  We gratefully acknowledge financial
support by Ohio University through the Rufus Putnam Visiting
Professorship, by the Volkswagen-Foundation, the RFBR (Russia),
and the US Department of Energy grant no.\ DE--FG02--87ER45334.
\mysection{References} 
\small \newcommand{\pbck}{\vspace{-1.1ex}}

\newpage

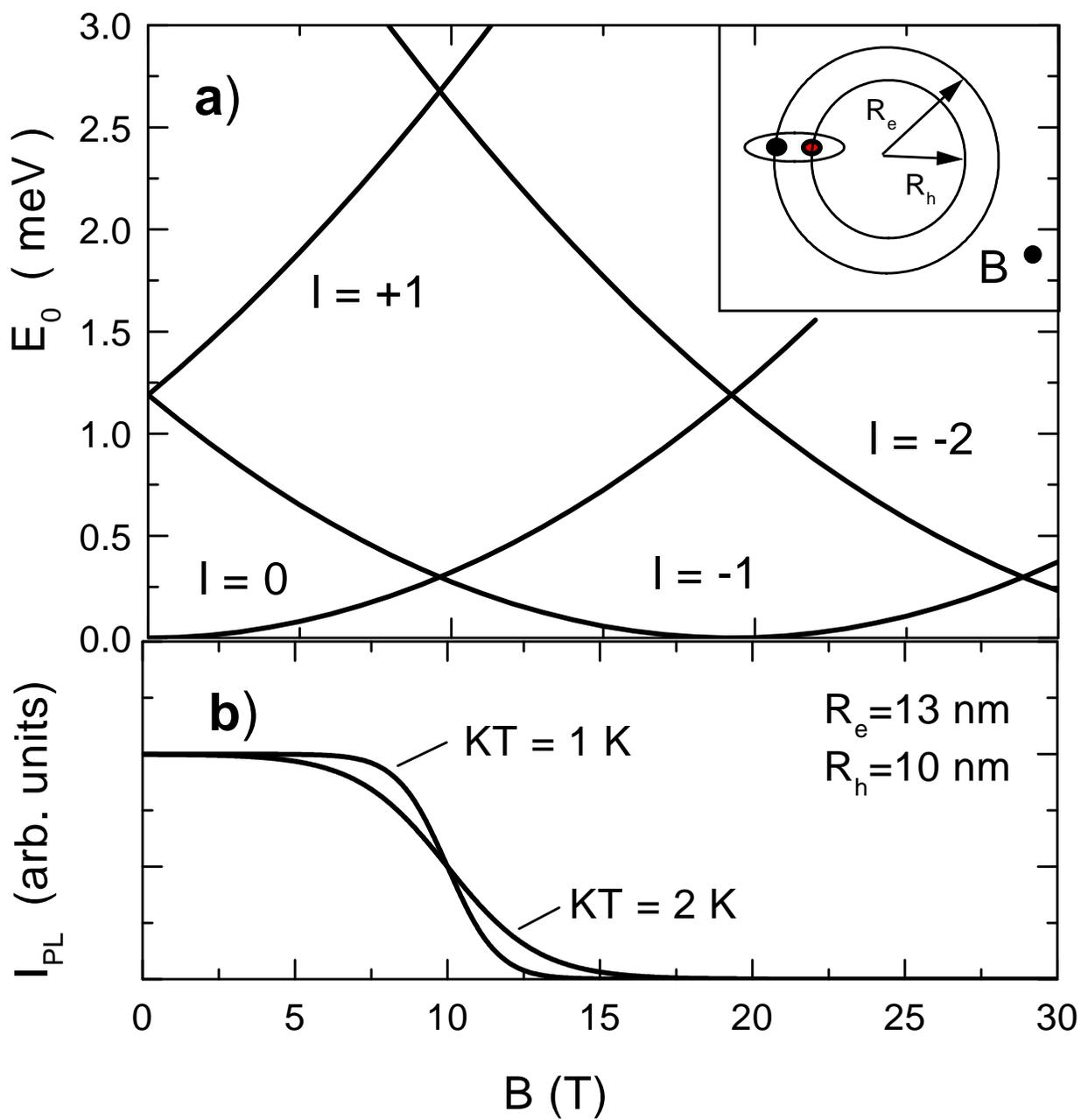
{\bf Figure captions } \\ \ \\ Fig.1 \ a) \ The energy spectrum
and PL intensity of excitons confined in a quantum ring as a
function of the normal magnetic field for the
strong-Coulomb-interaction limit; here, $m_e = 0.07 m_0$ and
$m_h=0.2 m_0$. Insert: a sketch of the quantum ring system. \\ \
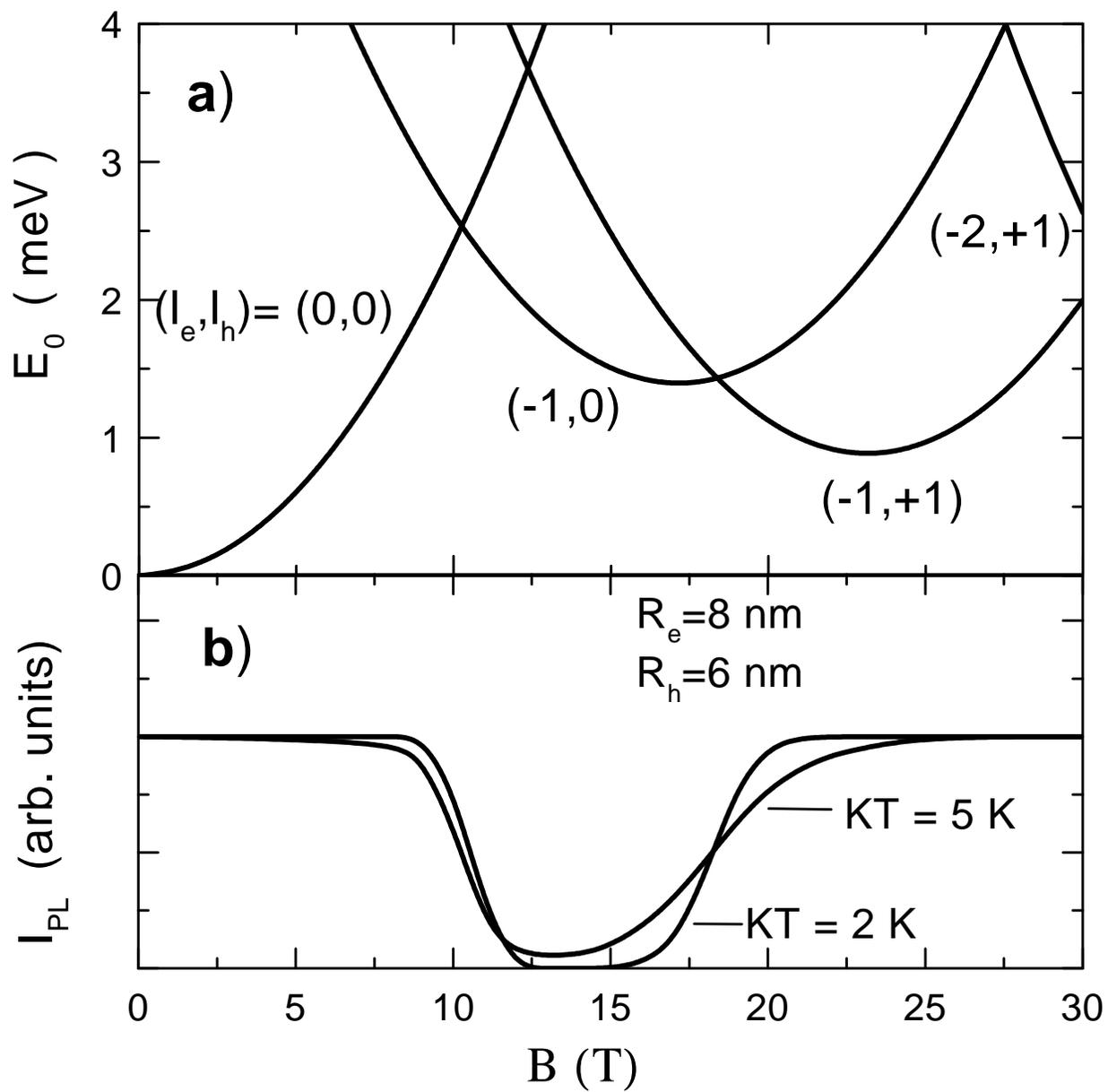
\\ Fig.2 \ b)\ The energy spectrum and PL intensity of excitons
confined in a quantum ring as a function of the normal magnetic
field in the limit of weak Coulomb interaction.


\begin{thebibliography}{31}
\bibitem{A}
The PL intensity includes also the overlap factor $A(B)^2=[\int
d\rho\rho f_e(\rho)f_h(\rho)]^2$, which decreases slowly with
magnetic field. For typical system parameters, the $B$-field
effect on $A$ is weak, since even at $B=30~T$ the magnetic length
is larger than $\Delta R=20 \sim 30$\AA. \pbck
\end{thebibliography}
\end{document}